\documentstyle[12pt]{article}

\newcommand{\be}{\begin{equation}}
\newcommand{\ee}{\end{equation}}
\newcommand{\ba}{\begin{eqnarray}}
\newcommand{\ea}{\end{eqnarray}}
\newcommand{\nn}{\nonumber}
\newcommand{\ie}{{\it{i.e.~}}}
\begin{document}
\def\pct#1{(see Fig. #1.)}

\begin{titlepage}
\hbox{\hskip 12cm ROM2F-97/32  \hfil}
\hbox{\hskip 12cm November, 1997 \hfil}
\vskip 1.4cm
\begin{center}  {\Large  \bf  Open \ Strings \ on the \ Neveu - Schwarz \ Pentabrane\footnote{This work
was supported in part by E.E.C. Grant CHRX-CT93-0340.}}
 
\vspace{1.8cm}
 
{\large \large Massimo Bianchi and Yassen S. Stanev\footnote{ On Leave from Institute for Nuclear
Research and Nuclear Energy, Bulgarian Academy of Sciences, BG-1784 Sofia, BULGARIA.}}
\vspace{0.5cm}

{ \sl Dipartimento di Fisica, \ \ Universit{\`a} di Roma \ ``Tor Vergata'' \\ I.N.F.N.\ - \
Sezione di Roma \ ``Tor Vergata'', \\ Via della Ricerca Scientifica , 1 \ \ 00133 \ Roma, \ \
ITALY}
\end{center}
\vskip 1.2cm

\abstract{ We analyze the propagation of open and unoriented strings on the  Neveu-Schwarz pentabrane
(N5-brane) along the lines of a similar analysis for the $SU(2)$ WZNW models. We discuss the two
classes of open descendants of the diagonal models and a series
of $Z_2$ projected models which exist only for even values of the level $k$ and correspond to
branes at D-type orbifold singularities. The resulting configurations of branes
and planes are T-dual to those relevant to the study of dualities in super Yang-Mills theories. The association of Chan-Paton factors to D-brane
multiplicities is possible in the semi-classical limit
$k\rightarrow\infty$,  but due to strong curvature effects is unclear for finite $k$. We show that the introduction of a magnetic field implies a twist of the $SU(2)$ current algebra in the open-string sector leading to spacetime supersymmetry breaking.}
\vfill
\end{titlepage}
\makeatletter
\@addtoreset{equation}{section}
\makeatother
\renewcommand{\theequation}{\thesection.\arabic{equation}}
\addtolength{\baselineskip}{0.3\baselineskip} 

\section{Introduction}

Starting from the initial proposal \cite{as} of interpreting open string theories as world-sheet
orbifolds of left-right symmetric closed string theories, open descendants of (Type II) models
have been  systematically constructed \cite{bs}. The worldsheet consistency conditions have been
further refined \cite{bps, min, fpss} and a large class of open descendants of rational Conformal
Field Theories (CFT) have been completely understood \cite{wzwa, wzwb}. It was not until the
advent of D-branes
\cite{cjp} that the above procedure, now termed ``orientifold", has received  so much attention
and interest. In the past two years D-brane techniques have lead to the (re)discovery
\cite{altri} of peculiar $N=(1,0)$ supersymmetric Type I vacuum configurations in $d=6$  with
various numbers of tensor multiplets and Chan-Paton (CP) symmetry breaking/enhancement initially
derived in \cite{bs} at rational points of their moduli spaces.  New Type I vacua in $d=6$ have
been derived as open descendants \cite{gep} of Gepner models and a chiral model in $d=4$ has
been found as the open descendant of the Type IIB theory on the Z-orbifold \cite{chir}. Other
Type I vacua in $d=4$ with or without D5-branes have been constructed
\cite{quattro} as ``orientifolds" of Type IIB compactifications on six-dimensional abelian
orbifolds.

More recently, D-branes \cite{hw, egk} and orientifolds \cite{ejs} have proved to be a powerful
alternative to the ``geometric engineering" \cite{bjpsv} of supersymmetric Yang-Mills (SYM)
theories. In this context the dynamics of D-branes
 and orientifold planes (O-planes) in the background of NS pentabranes (N5-branes) \cite{chs}
seems to provide a geometrical interpretation of dualities in some SYM theories
\cite{aps}. A unifying picture emerged from the proposal \cite{witmt} that the relevant
configurations of N5-branes and D-branes be interpreted as a single M5-brane (M-theory 5-brane)
wrapped around a Riemann surface.
Contrary to the vacuum configurations with D-branes invading all the non-compact spacetime
dimensions, the configurations of D-branes and O-planes relevant to the study of SYM theories
allow for the R-R charge  to leak out at infinity and the tadpole consistency conditions are not to be imposed \cite{ib}. Moreover, in the cases that we discuss there are no massless closed-string states and it is meaningless to 
cancel massive tadpoles, even for branes in compact spaces.

In this paper we apply the by now standard procedure for open-string descendants of rational
conformal field theories (RCFT) to the study of open and unoriented string propagation in the
presence of N5-branes. After reviewing some known facts about N5-branes \cite{chs}, we determine
the correct parametrization of the spectrum of the open and unoriented strings in the cases of
factorized diagonal models and for a series of nonfactorized
$Z_2$ projected models that exist only for even level $k$ \cite{afk}. The scaling of the various
amplitudes (Annulus, M\"obius strip, Klein bottle) for large $k$ allows one to identify the relevant configuration of D-branes and O-planes
in the N5-brane background. We argue that tadpole cancellation needs not to be imposed and
discuss the association of CP charges to D-brane multiplicities in the limit
$k\rightarrow\infty$. We also argue that due to strong curvature effects the distinction between
different kinds of branes becomes unclear for finite $k$. Some useful formulae for the
open-descendants of the $SU(2)$ WZNW models \cite{wzwa, wzwb} are collected in the Appendix.
Finally we show that the addition of a magnetic field induces a twisting of the
$SU(2)$ current algebra in the open sectors that implies a breaking of spacetime supersymmetry
\cite{acny}.

Similar issues have been considered recently in \cite{fgp} but we find their analysis incomplete
and to some extent inconsistent with the non-abelian structure of the fusion algebra for $k >
1$. 

\section{N5-branes, D-branes and O-planes}

String solitons with NS-NS magnetic charge correspond to extended objects with  a
$5+1$-dimensional worldvolume, \ie pentabranes or briefly N5-branes \cite{chs}. For Type II
superstrings, the background is completely characterized by setting to zero all the R-R fields
and taking
\ba   ds^2 &=& \eta_{\mu\nu} dx^{\mu} dx^{\nu} + e^{-2\phi} (dr^2 + r^2 ds^2_3) 
\nn
\\ e^{-2\phi} &=&  e^{-2\phi_o}  (1 + {k\over r^2}) 
\\ 
\label{penta}
  H &=& dB = * de^{-2\phi} = - k d\Omega_3 \quad ,
\nn
\ea where the indices $\mu,\nu = 0,1, \dots ,5$ are tangent to the N5-brane, while  $ds^2_3$ and
$d\Omega_3$ are the line and volume elements on $S^3$, respectively. The geometry of the space
transverse to the N5-brane is that of a semiwormhole with the size of the throat fixed by the
axionic charge $k$ (the number of coincident N5-branes). In the throat region $(r\rightarrow 0)$ the
dilaton diverges and the N5-brane background admits an exact CFT description as
the tensor product of a $SU(2)$ WZNW model at level $k$ times a Feigin-Fuchs boson $X_4$ with
background charge $Q=\sqrt{2/(k+2)}$. After including the fermionic partners $\{\psi^i,
\psi^4\}$, the world-sheet theory gains an extended  $N=(4,4)$ superconformal symmetry which
guarantees the absence of both perturbative and non-perturbative corrections in $\alpha^{\prime}$
\cite{chs}. The   modular invariant spectrum of the closed string
excitations around the semi-wormhole background for even values of $k$ has been worked out in
\cite{afk}, where also other classes of 4-d backgrounds with exact $N=(4,4)$ superconformal
symmetry have been constructed. 

Many other exact 4-d backgrounds (corresponding to generalized hyper-K\"ahler manifolds) and
their Buscher T-duals \cite{bus} have been analyzed in
\cite{kkl} in relation to non-compact Calabi-Yau manifolds and axionic instantons. Stringy ALE
instantons were thoroughly analyzed in \cite{ale}. More recently, string dualities in $d=6$ have
been given support by the observation that Type IIA (B) in the presence of  $k$ coincident
N5-branes is equivalent (Buscher T-dual) to Type IIB (A) superstring on an ALE space
($R^4/\Gamma_k$)
\cite{ov} at vanishing B-field \cite{asp}. The crucial observation is that the 6-d effective
field theory of the symmetric N5-branes displays $N=(2,0)$, respectively $N=(1,1)$,
supersymmetry for the Type IIA, respectively Type IIB, superstring \cite{chs}. For the Type IIB
N5-brane one expects an
$N=(1,1)$ vector multiplet whose four scalar components are the collective coordinates for the
translation of the N5-brane in the transverse 4-d space. This fits with the conjectured $SL(2,Z)$
U-duality of the 10-d Type IIB superstring which relates the N5-brane to the D5-branes,  the
world-volume degrees of freedom of the latter being massless open-string excitations in an
$N=(1,1)$ vector multiplet. On the contrary, the Type IIA N5-brane requires an $N=(2,0)$ tensor
multiplet with 5 scalars, that are very suggestive of an 11-d interpretation in terms of
M-theory. Indeed the Type IIA N5-branes are conjectured to arise from M5-branes  \cite{witdy}.  M5-branes that
are wrapped  around the eleventh dimension give rise to D4-branes.   

Using U-duality one can show that to a given configuration of parallel N5-branes one can add
D4-branes terminating on them and (compatibly with the surviving supersymmetry) a collection of
transverse D6-branes \cite{egk, ejs}. Denoting by $x^0, x^1, x^2, x^3, x^4, x^5$ the coordinates
tangent to the  N5-brane world-volume, the D4-brane worldvolume may be taken along the directions
$x^0, x^1, x^2, x^3, x^6$ (the last direction is ``compactified" either on a  segment or on a
circle), while the D6-brane worldvolume may be taken  along the directions
$x^0, x^1, x^2, x^3, x^7, x^8, x^9$. Denoting by $N_4$ and $N_6$ the number of D4-branes and 
D6-brane respectively, the effective gauge theory \cite{witbp, dm} on the non-compact directions
of the world-volume of the D4-brane ($x^0, x^1, x^2, x^3$) is
$N=2$ SYM with gauge group contained in $U(N_4)$ and a ``flavor" symmetry contained in $U(N_6)$  
\cite{egk}. One can also add Op-planes parallel to the Dp-branes and break the gauge/flavor
symmetry to orthogonal or symplectic groups \cite{ejs}. Performing a T-duality on the
directions  $x^4$ and
$x^5$ the  D4- (D6-) branes turn into D6- (D8-) branes \cite{cjp}. 

The microscopic string excitations of BPS configuration of D-branes and O-planes in the
background of $k$ coincident N5-branes can be explicitly determined as open-string descendants
of the rational CFT that describes the throat region.  After discussing the case of N5-branes in
a priori flat spacetime we will also discuss the open-descendants of the configurations studied
in \cite{afk, kk}, that we interpret as  N5-branes at D-type singularities \cite{sw, ib}.
Separating the N5-branes or rotating some of them  \cite{egk, ejs} in order to obtain $N=1$
configurations seems to be out of the reach of our simple  CFT analysis so far and remains a
challenge for future work on the subject.

\section{From Spheres to Semi-Wormholes}

After an anomalous chiral transformation, the fermionic partners $\psi^i$ of the bosonic 
coordinates $X^i$ decouple from the $SU(2)$ currents $J^i$. The prize one has to pay is a
finite renormalization of the level $k\rightarrow k-2$ \cite{chs}. The only interacting degrees
of freedom for the N5-brane CFT are the bosonic coordinates of the $SU(2)$ group manifold,
$S^3$. Open string propagation on $S^3$ has been considered in connection to 2-d charged black
holes
\cite{hor}. The problem was thoroughly addressed along the lines of \cite{bs} and completely
solved for the $A$, $D_{even}$ and $E$ series in \cite{wzwa} and for the $D_{odd}$ series  in
\cite{wzwb}.  Some relevant formulae are collected in the Appendix.   The central charge of the
Virasoro algebra of the $SU(2)$ WZNW models  at level $k$ is
$c = 3k/(k+2)$.  The conformal weights of the integrable unitary representations are 
\be h_I^{(k)} = {I(I+1)\over k+2}
\label{wieght}
\ee with isospin $I$ in the range $I=0,..,k/2$. The generalized character formula is given by
\cite{kac}
\ba
\chi_I^{(k)}(\tau,z,u) &=& Tr_{{\cal H}_I^{(k)}} q^{L_o-{c\over 24}} e^{2\pi i z J_o^{(3)}} = \nn
\\      &=&  e^{2\pi i k u} { {q^{h_I^{(k)}-{c(k)\over 24}} \sum_{n} q^{(k+2)n^2+(2I+1)n}
sin[\pi z (2I+1+2 n(k+2))]}
\over {sin (\pi z) \prod_{n=1}^{\infty} (1-q^n) (1-e^{2\pi i z} q^n) (1-e^{-2\pi i z}q^n)} } 
\label{character}
\ea

The unoriented projection can be chosen to preserve the diagonal $SU(2)$ subalgebra of the
$SU(2)_L \times SU(2)_R$ current algebra. Corresponding to the two geometrical involutions on
the $SU(2)$ group manifold
$g\rightarrow -g^{-1}$ and $g\rightarrow g^{-1}$, there are two different unoriented projections 
(Klein bottle amplitudes) of the parent torus partition function.  Notice that the
two involutions have the same action on the integer isospin representations. Thus  the two projections differ only  for the cases of diagonal ($A$)
and $D_{odd}$ modular invariants. On the half-integer isospin representations the first one
corresponds to keeping  the singlets of the diagonal $SU(2)$, while the second one corresponds to removing
them from the spectrum.

A similar analysis can be carried over directly to the open-descendants of the Type IIA
superstring in the background of $k$ N5-branes. One simply has to combine the $SU(2)$
characters  (\ref{character}) with the contributions of the FF boson $X^4$, the flat bosonic
coordinates $X^\mu$ and the fermions
$\{\psi^\mu, \psi^i, \psi^4 \}$. The latter contribution is easily expressed in terms of
$\theta$-functions or better, for the purpose of reading off the spectrum, in terms of  the
characters of the four integrable representations of $SO(2n)$ at level one $\{O_{2n}, V_{2n},
S_{2n}, C_{2n}\}$. The FF boson and each non-compact coordinate give the standard contribution
${(\sqrt{\tau_2} |\eta|^2)}^{-1}$. The discrete states of the FF boson are a set of zero measure
and do not contribute to the partition function \cite{afk, kk}. They however play a crucial role
in the computation of correlation functions where they act as screening operators for the
background charge. It would be interesting to see whether they can be related
to the
collective coordinates of the N5-brane.

For N5-branes embedded in a flat space-time the torus partition function for the Type IIA
superstring assumes a factorized form
\footnote{Were it not for the addition of D-branes and O-planes, one would have had to truncate
the spectrum as in \cite{afk, kk} via a generalized GOS projection that correlate the internal
isospin to one of the two transverse Lorentz spins.}
\be {\cal T} = (V_8-S_8) (\bar{V_8} -\bar{C_8}) \sum_{ab} I_{ab} \chi_a \bar\chi_b
\label{torus}
\ee where $I_{ab}$ is one of the $A-D-E$  modular invariant combination of
$SU(2)$ characters\footnote{ The three exceptional cases corresponding to the
$E$ modular invariants do not have however  a clear geometrical interpretation.} that we label
by the dimension of the corresponding representation $a=2I+1$. The 
unoriented projection
is similar to the pure
$SU(2)$ case (see the Appendix). Note that the partition function (\ref{torus}) is effectively
left-right symmetric, since  the
 left-right interchange in the $SU(2)$ factor, corresponding to 
$g\rightarrow \pm g^{-1}$, has to be  combined with a flip of the chirality of the spinors $S_8
\leftrightarrow \bar{C_8}$ \cite{ejs}.  This is consistent with the introduction of Dp-branes
with p even. 

Since the volume of the throat of the wormhole is quantized in units of $k$, one
can use the overall factor of $k$ to trace the scaling of the open and unoriented amplitudes
with the volume. Let us focus on the two descendants of the $A$-series.  One can analyze the
large $k$ behavior of the partition functions given in the Appendix, keeping only the states
that become massless in this limit 
 (that is the states with $a < \sqrt{k}$). The Klein bottle expression for real CP charges is
independent of
$k$ for large $k$ plus subleading terms. This we interpret as an indication that the
unorientifold introduces D6-branes and O6-planes in this case. 
The resulting CP group is  a product of   orthogonal  and symplectic factors,
namely $\prod_{a \ odd} SO(n_a) \times \prod_{b \ even} Sp(n_b)$ or 
$\prod_{a \ odd} Sp(n_a) \times \prod_{b \ even} SO(n_b)$
depending on the overall sign  in (\ref{areald}).
For complex CP charges, one finds
amplitudes that up to subleading terms behave as $k$ for large $k$. This we interpret as an
indication that the unorientifold introduces D8-branes and O8-planes in this case.   The resulting CP group is a product of unitary factors $U(n_a)$
times (for even level $k$ only) an $SO(n_{\rho})$ or $Sp(n_{\rho})$ 
(where $\rho = k/2+1$) factor.
Depending on the choice of the nonvanishing CP charges,
the annulus partition functions of both classes of models
either grow linearly or are independent of $k$.

If some of the non-compact directions $x^1 \dots x^5$ are compactified,  by 
T-duality transformations one can generate from the D6- and D8-branes any configuration of Dp-
and D(p+2)-branes with p in the range $1 \dots 6$, where even(odd)  p correspond to the
descendants of the Type IIA(B) superstring.

\section{Pentabranes at orbifold singularities}

Another possibility is to put the N5-branes at orbifold singularities \cite{ib}.  In the CFT
approach, a simple way to achieve this for even values of the level $k$ is to entangle the
fermion boundary conditions with the  $Z_2$ action on the $SU(2)$ currents that generates the
D-type modular invariant from the diagonal A-type modular invariant. The resulting configuration
admits a geometrical interpretation in terms of  N5-branes at D-type orbifold singularities of
$R^4$. For the Type II superstring the spectrum is encoded  in the modular invariant torus
partition  function of \cite{afk},  that can be conveniently re-expressed in terms of the
``supercharacters" \cite{bs}
\ba   Q_o &=& V_4 O_4 - S_4 S_4 \quad , \quad Q_v = O_4 V_4 - C_4 C_4 \nn \\ Q_s &=& O_4 C_4 -
S_4 O_4 \quad , \quad Q_c = V_4 S_4 - C_4 V_4 
\label{superQ}
\ea  The introduction of D-branes breaks the accidental $SO(4)$ symmetry to the $SO(3)\equiv SU(2)$
automorphism of the spacetime superalgebra. The characters of the internal $SO(4)$ would then
break according to 
\be O_4 = O_3 O_1 + V_3 V_1 ,  \qquad V_4 = V_3 O_1 + O_3 V_1 ,  \qquad S_4 = \sigma_3 \sigma_1
= C_4
\label{rottura}
\ee  where $\{O_1, V_1, \sigma_1\}$ are the characters of the Ising model associated to the FF
fermion $\psi^4$. After resolving the ambiguity in the modular transformation matrix S one
however recovers the above characters $\{Q_o, Q_v, Q_s, Q_c\}$.  Neglecting the (modular
invariant) contributions of the bosonic coordinates as well as of the FF boson, the
closed-string untwisted sector reads
\ba  {\cal T}_u &=& {1\over 2} \left (|Q_o + Q_v|^2 (\sum_{odd \ a} |\chi_{a}|^2 + \sum_{even \
a } |\chi_{a}|^2) \right. 
\nn
\\
 &+& \left. |Q_o - Q_v|^2 (\sum_{odd \ a} |\chi_{a}|^2 - \sum_{even \ a} |\chi_{a}|^2)\right)
\label{untwist}
\ea where $a$ is the shifted weight $a=2I+1 \ (\leq k+1)$. Thus odd/even $a$ corresponds to
integer/halfinteger isospin $I$. A modular S transformation displays the twisted sector
\ba {\cal T}_t &=& {1\over 2} \left( |Q_s + Q_c|^2 (\sum_{odd \ a} \chi_{a}
\bar
\chi_{k+2-a}  + \sum_{even \ a} \chi_{a} \bar
\chi_{k+2-a}) \right.
\nn
\\ &+& \left. (-)^{k/2} |Q_s - Q_c|^2 (\sum_{odd \ a} \chi_{a} \bar
\chi_{k+2-a}  - \sum_{even \ a} \chi_{a} \bar
\chi_{k+2-a}) \right)
\label{twisted}
\ea The spectrum consists of the $N=(2,0)$ supergravity multiplet and five
$N=(2,0)$ tensor multiplets with mass shifted from zero by the non-trivial dilaton background
\cite{afk}, \ie $m^2=Q^2/8= 1/4(k+2)$. The twisted sector gives rise only to massive excitations
except for the case
$k=2$, where however the semiclassical analysis is inappropriate and the model has to be
interpreted in terms of non-critical $N=2$ strings \cite{afk}. For the sake of a geometrical
interpretation we will stick to the semiclassical regime of very large $k$. 

There is a striking similarity between the modular invariant torus amplitude for the propagation
of Type II superstrings in the wormhole geometry and the
$SU(2)_{4p+2}$ WZW model  with a $D_{odd}$  type modular invariant. In particular only the fields
corresponding  to the characters $Q_o \chi_{a}$, $Q_v \chi_{a}$, with odd $a$, $Q_s \chi_{\rho}$
and $Q_c \chi_{\rho}$ may flow in the ``tube" channel and can enter in the direct channel Klein
bottle amplitudes
\be K_{\pm} = (Q_o + Q_v) \sum_{odd \ a=1}^{k+1} \chi_{a} \pm (Q_s +Q_c) \chi_{\rho}
\label{ykd}
\ee 
where $\rho=k/2+1$.
The overall sign of the projection in the unoriented closed-string spectrum is the only
freedom left by imposing the consistency conditions. 

In order to determine the correct CP charge assignments one has to determine first the relevant
boundary states.  The transverse channel annulus amplitude can then be written in the form 
\be
\tilde A = \sum_j \tilde \chi_j \left[ \sum_{\alpha}  B_j^{(\alpha)} n_{\alpha} {C^{\alpha}_b
\over
\sqrt{C^{j}_v}} \right]^2 ,
\label{atilb}
\ee where $C^{\alpha}_b$ and $C^{j}_v$ and   are the normalizations of the boundary and bulk
2-point functions, while $B_j^{(\alpha)}$ are the reflection coefficients 
 of the fields  in the 
$j^{th}$ sector of the spectrum from the $\alpha^{th}$ type of boundary. With an appropriate
choice of the normalizations $C_b$ and $C_v$  the reflection coefficients 
 $B_i^{(\alpha)}$ satisfy
\be   B_i^{(\alpha)}B_j^{(\alpha)} = \sum_k \sigma_{ij}{}^k N_{ij}{}^k B_k^{(\alpha)}
\label{bsqb}
\ee where $N_{ij}{}^k$ are the fusion rule coefficients and $\sigma_{ij}{}^k$ are just signs 
\cite{wzwb}.  In the case at hand, the non trivial part of the CFT comes from the coupling 
 of the WZW model and the supercharacters $Q_x$, with $x=o,v,s,c$. Thus  in (\ref{atilb}) $\tilde
\chi_j = Q_x \chi_a$ and the index $j$ corresponds to  the  pair $(x,a)$. The signs $\sigma$
are symmetric in all pairs of indices and are given by 
\be
\sigma_{ax,by,dz}= \left\{  { (-1)^{d-1 \over 2} \quad {\rm if \ both \ x \ and \ y \ are \
equal
\ to \ s
\ or
\ c}
\atop 
\ +1 \qquad \quad { \rm else} \hfill }
\right .
\label{signs1}
\ee Note that for levels $k=4p+2$ there is an equivalent definition independent of the values of
$x$,
$y$ and $z$ 
\be
\sigma_{ax,by,dz}= \sigma_{a,b,d} =  \left\{  { (-1)^{d-1 \over 2} \quad {\rm if \ both }\ a
\ {\rm and} \ b \ {\rm are
\ even} \atop 
\ +1 \qquad \quad { \rm else} \hfill }
\right .
\label{signs2}
\ee thus for these values of $k$ one can directly use the results of \cite{wzwb} for the $SU(2)$
model. For levels $k= 4p$ however one must solve explicitly  the system  (\ref{bsqb}) for the
reflection coefficients. The construction is simplified by the observation  that the subset of
fields of integer isospin is common both to the diagonal ( $A$ ) and  the $D$ model and the
corresponding reflection coefficients coincide, so can be read from (\ref{arealt}). The
remaining reflection coefficients are then easily determined from  (\ref{bsqb},\ref{signs1}).

After a modular transformation $S$ from the direct Klein bottle projections (\ref{ykd}) we obtain for 
the transverse Klein bottle amplitudes
\ba
\tilde K_{\pm}  =   \sqrt{{ 1 \over \rho}}  \sum_{a \ odd } {\chi_a \over sin({a \pi \over
k+2})} &&
\left\{ Q_o  
\left [ \sqrt{2} (-1)^{a^2-1 \over 8} sin \left(  {a \pi (\rho \pm 1) \over 2(k+2)} \right)
\right ]^2 +
\right.
\nn
\\ && \quad \left. Q_v  
\left [ \sqrt{2} (-1)^{a^2-1 \over 8} sin \left( {a \pi (\rho \mp 1) \over 2(k+2)}\right) \right
]^2 
\right\}
\label{ykt}
\ea Note that for $k=4p+2$ this expression is a linear combination of the real and complex  Klein
bottle amplitudes in the pure $SU(2)$ WZW model (\ref{kdodrealt},\ref{kdodcomplext}). In fact
this is valid  for all the amplitudes below, thus the following expressions are an alternative
representation of the CP charge assignments obtained in \cite{wzwb}.

The corresponding transverse channel annulus amplitudes are 
\ba && \tilde A_{\pm} = \sqrt{{ 1 \over \rho}} \sum_{a \ odd } {\chi_a \over sin({a \pi \over
k+2})}
\times
\nn
\\ && \left\{
 Q_o 
\left [ \sqrt{2} \sum_{\alpha=1}^{k/2} (n_{\alpha}^{(1)}+n_{\alpha}^{(2)})  sin \left({\alpha a
\pi  \over k+2} \right) + {(-1)^{a^2-1 \over 8} \over \sqrt{2}} (n_{\rho}^{(1)} + \tilde
n_{\rho}^{(1)} + n_{\rho}^{(2)} + \tilde n_{\rho}^{(2)} ) \right ]^2 +
\right .
\nn
\\ && \left . Q_v 
\left [ \sqrt{2} \sum_{\alpha=1}^{k/2} (n_{\alpha}^{(1)}-n_{\alpha}^{(2)})  sin \left ({\alpha a
\pi  \over k+2} \right) + {(-1)^{a^2-1 \over 8} \over \sqrt{2}} (n_{\rho}^{(1)} + \tilde
n_{\rho}^{(1)} - n_{\rho}^{(2)} - \tilde n_{\rho}^{(2)} ) \right ]^2 
\right\} \pm
\nn
\\ && \pm (-1)^{\rho} \sqrt{\rho} \chi_{\rho} 
\left\{ Q_s  \left[{1 \over \sqrt{2}}  (n_{\rho}^{(1)} - \tilde n_{\rho}^{(1)} + n_{\rho}^{(2)} -
\tilde n_{\rho}^{(2)} ) \right ]^2 +
\right.
\nn
\\ && \qquad \qquad \qquad \left. Q_c  \left[{1 \over \sqrt{2}}  (n_{\rho}^{(1)} - \tilde
n_{\rho}^{(1)} - n_{\rho}^{(2)} + \tilde n_{\rho}^{(2)} ) \right ]^2 
\right\}
\label{yat}
\ea where $n^{(1)}$ and $n^{(2)}$  are two sets of CP charges. The charges $n_{\alpha}$  
correspond to boundary states that are linear combinations of the ones in the diagonal case, 
while $n_{\rho}$ and $\tilde n_{\rho}$ correspond to a splitting of one diagonal boundary state
thus giving rise to multiplicities that have no counterpart in the closed sector.  After a
modular $S$ transformation we obtain for the  Annulus partition function
\ba A_{\pm} &=& (Q_o+Q_v) \sum_a   \chi_a   \left [ \sum_{\alpha, \beta = 1}^{k/2} 
  A_{a \alpha \beta}  (n_{\alpha}^{(1)} n_{\beta }^{(1)} + n_{\alpha}^{(2)} n_{\beta }^{(2)} ) +
\right.
\nn
\\ && \qquad \qquad \qquad \qquad \left.
 \sum_{ \beta = 1}^{k/2} 
  A_{a \rho \beta}  (n_{\rho}^{(1)} n_{\beta }^{(1)} + 
\tilde n_{\rho}^{(1)}  n_{\beta }^{(1)} +
 n_{\rho}^{(2)} n_{\beta }^{(2)} +
\tilde  n_{\rho}^{(2)}  n_{\beta }^{(2)}) \right ] +
\nn
\\ && (Q_s+Q_c) \sum_a   \chi_a   \left [ \sum_{\alpha, \beta = 1}^{k/2} 
  A_{a \alpha \beta}  (2 n_{\alpha}^{(1)} n_{\beta }^{(2)} ) +
\right.
\nn
\\ && \qquad \qquad \qquad \qquad \left.
 \sum_{ \beta = 1}^{k/2} 
  A_{a \rho \beta}  (n_{\rho}^{(1)} n_{\beta }^{(2)} + 
\tilde n_{\rho}^{(1)}  n_{\beta }^{(2)} +
 n_{\rho}^{(2)} n_{\beta }^{(1)} +
\tilde  n_{\rho}^{(2)}  n_{\beta }^{(1)}) \right ] +
\nn 
\\ && Q_o \sum_{a \ odd}   \chi_a  \left[  {1 \pm \varepsilon^{k,a} \over 2} \left (
(n_{\rho}^{(1)})^2+( \tilde n_{\rho}^{(1)})^2+ (n_{\rho}^{(2)})^2+( \tilde n_{\rho}^{(2)})^2
\right ) + \right.
\nn
\\ && \qquad \qquad \quad  \left. {1 \mp \varepsilon^{k,a} \over 2} \left ( 2 n_{\rho}^{(1)}
\tilde n_{\rho}^{(1)}+ 2 n_{\rho}^{(2)}  \tilde n_{\rho}^{(2)} \right ) \right ] +
\\ && Q_v \sum_{a \ odd}   \chi_a  \left[  {1 \mp \varepsilon^{k,a} \over 2} \left (
(n_{\rho}^{(1)})^2+( \tilde n_{\rho}^{(1)})^2+ (n_{\rho}^{(2)})^2+( \tilde n_{\rho}^{(2)})^2
\right ) + \right.
\nn
\\ && \qquad \qquad \quad  \left. {1 \pm \varepsilon^{k,a} \over 2} \left ( 2 n_{\rho}^{(1)}
\tilde n_{\rho}^{(1)}+ 2 n_{\rho}^{(2)}  \tilde n_{\rho}^{(2)} \right ) \right ] +
\nn  
\\ && Q_s \sum_{a \ odd}   \chi_a  \left[  {1 \mp \varepsilon^{k,a} \over 2} \left ( 2
n_{\rho}^{(1)}  n_{\rho}^{(2)}+ 2 \tilde n_{\rho}^{(1)} \tilde n_{\rho}^{(2)} \right ) +  {1 \pm
\varepsilon^{k,a} \over 2} \left ( 2 n_{\rho}^{(1)} \tilde n_{\rho}^{(2)}+ 2 \tilde
n_{\rho}^{(1)}   n_{\rho}^{(2)} \right ) \right ] +
\nn  
\\ && Q_c \sum_{a \ odd}   \chi_a  \left[  {1 \pm \varepsilon^{k,a} \over 2} \left ( 2
n_{\rho}^{(1)}  n_{\rho}^{(2)}+ 2 \tilde n_{\rho}^{(1)} \tilde n_{\rho}^{(2)} \right ) +  {1 \mp
\varepsilon^{k,a} \over 2} \left ( 2 n_{\rho}^{(1)} \tilde n_{\rho}^{(2)}+ 2 \tilde
n_{\rho}^{(1)}   n_{\rho}^{(2)} \right ) \right ] 
\qquad
\nn
\label{yad}
\ea where $\varepsilon^{k,a} = (-1)^{{k \over 2 } + 1 + {a - 1 \over 2 }} $, and
\be A_{a \alpha \beta} = 2 \sum_{b \ odd}  {S_{ab} S_{\alpha b } S_{\beta b} \over S_{1b}} \quad
.
\label{ynf}
\ee  For the $+(-)$ case and even(odd)  $k/2$  one has to use pairs of complex charges 
$\tilde n_{\rho}^{(i)} = \bar n_{\rho}^{(i)} \ ( =  n_{\rho}^{(i)} )$, while  the other
parametrization is real. Note that the coefficients $A_{a  \alpha  \beta}$ are {\it NOT}
proportional to the standard
$SU(2)$ fusion rules  since the sum goes only over the odd values of $b$. Nevertheless all $A_{a
\alpha \beta}$ take  nonnegative  integer  values (0,1 and 2 to be precise). Eq. (\ref{ynf})
gives the simplest (and best understood) example of the general construction of
\cite{schwf}. There are again two distinct limits for large $k$ of these annulus amplitudes. This
can be  demonstrated by considering the leading terms in the amplitude proportional to
$n_{\alpha}^2$. In particular the limit of the contributions  with $\alpha $ small with respect
to
$\sqrt{k}$ is independent of $k$ (since {\it e.g.} $A_{a 1 1 }=\delta_{a1}$), while the limit of the
contributions with $\alpha$ close to  $k/2$ grows linearly with $k$ (since all $A_{a  {k
\over 2} { k \over 2}}$ are nonvanishing). This we again interpret as an indication of the
presence of D6-branes in the former and of D8-branes in the latter case. It is important to note
that in all cases the Klein bottle amplitudes grow  linearly with $k$ in the limit $k
\rightarrow \infty$. This we interpret as  an indication that there are always O8-planes
present. The subleading terms  that  correspond to O6-planes are not easy to trace.   In
the cases when the annulus and Klein bottle amplitudes  have different  large
$k$ limits it is impossible (and in fact not necessary) to  impose a cancellation of the
(massive!) tadpoles.

Since there are no tadpole conditions, the transverse M\"obius amplitude  is only defined up to
an overall sign and is obtained  in the standard way from the transverse annulus and transverse
Klein  bottle amplitudes where  we have made explicit also the relevant signs of the reflection
coefficients. After a modular 
$P^{\dagger}$  ($=P$ in the case at hand) transformation for the direct channel we get 
\ba M_{\pm} =  \left\{ \hat Q_o \sum_{a \ odd}  \hat \chi_a
\left [\sum_{\alpha=1}^{k/2} (n_{\alpha}^{(1)}-n_{\alpha}^{(2)}) M^{\mp}_{a \alpha} +
(n_{\rho}^{(1)} + \tilde n_{\rho}^{(1)} - n_{\rho}^{(2)} - \tilde n_{\rho}^{(2)}) {1 \pm
\varepsilon^{k,a} \over 2} \right] + \right .
\nn  
\\
\left .
\hat Q_v \sum_{a \ odd} \hat \chi_a
\left [\sum_{\alpha=1}^{k/2}  (n_{\alpha}^{(1)}+n_{\alpha}^{(2)}) M^{\pm}_{a \alpha} +
(n_{\rho}^{(1)} + \tilde n_{\rho}^{(1)} + n_{\rho}^{(2)} + \tilde n_{\rho}^{(2)}) {1 \mp
\varepsilon^{k,a} \over 2} \right] \right \}
\label{ymd}
\ea where the coefficients $M^{\pm}_{a \alpha }$ are given by
\be
 M^{\pm}_{a \alpha } =  {2  \over \sqrt{ \rho } }\sum_{b \ odd}   (-1)^{{b^2-1 \over 8 }}
sin\left({b(\rho \pm 1) \pi \over 2(k+2)}\right)
 { S_{\alpha b } P_{a b} 
\over S_{1b}} 
\label{ymhat}
\ee

It is instructive to verify that $M_{\pm}$ give a consistent symmetrization of the above annulus
partition functions. In particular one can demonstrate that :

- $A_{a \alpha \alpha} = 0 $ for all even $a$ and all $\alpha \leq k/2$;

- $A_{a \alpha \alpha} = M^{\pm}_{a \alpha} $ (mod 2)  for all odd
 $a$ and all $\alpha \leq k/2$.

The resulting open-string spectrum in principle contains tachyons  coming from the combinations
of $Q_s$ with $\chi_a$ with small $a$, 
\ie $a<\sqrt{k+2}$. In order to remove them it is sufficient to put to zero either one of the
two sets of CP
 multiplicities, say $n^{(2)}_\alpha=0$ for all $\alpha$. An explicit analysis of some small $k$
examples indicates however that this  choice is not necessary. Of particular interest are the
models with height equal to a
 square integer $k+2 = \ell^2$ for even $\ell \geq 4$, since their open spectra generically 
contain also massless states\footnote{As already noted the case $k=2$, which corresponds to $\ell=2$, is degenerate and
 contains massless states also in the closed spectrum.}. Indeed the conformal weight of the lowest lying states in 
$Q_s \chi_\ell$ is exactly $1/2$.  As an example let us consider the simplest such case
corresponding to $k=14$, where there  are 9 inequivalent choices of the non-vanishing CP
multiplicities that remove the tachyons  from the spectrum while keeping the massless open
string state. We shall present only one of these solutions that generalizes for all  (even)
$\ell$, namely if one chooses all $n^{(1)}_{\alpha}=0$ with $\alpha=1,\dots,k/2$, but
$n^{(1)}_\rho$  and
$\tilde n^{(1)}_\rho$ are nonvanishing, to remove the tachyons it is  sufficient to  put to zero
only the charges $n^{(2)}_\alpha$ with $\alpha \geq \ell(\ell-2)/2 + 2$ as well  as 
$n^{(2)}_\rho$ and $\tilde n^{(2)}_\rho$. 
It is not clear whether the presence of massless chiral fermions in the 
open-string spectrum implies any anomaly cancellation
condition in a background where both the gravity and the vector multiplets 
are massive for any finite $k$.

The resulting CP group is a product of 
orthogonal  and symplectic factors times (only in the +(-) case and even(odd) $k/2$) a unitary $U(n_\rho)$ factor. 
Due to the absence of tadpole conditions (only massive states appear in the transverse spectrum) 
the values of 
the non-vanishing CP multiplicities remain totally undetermined. Moreover
their correspondence with different Dp-branes multiplicities is far from
being obvious for finite values of the level $k$, due to strong curvature effects.

As a side remark, notice that
our derivation of the partition functions started from the transverse channel expressions and 
required a detailed knowledge of the $2d$ structure constants to determine the signs 
(\ref{signs1}). It  should be stressed however that there is an alternative derivation of these
partition  functions that uses only the fusion rules and the modular transformation matrices. 
Namely, one has  to solve the polynomial relations 
\cite{wzwb} for the integer valued  direct channel 
expressions $A_{a \alpha \beta}$,
$M_{a \alpha}$ and $K_a$, the fusion rules 
$N_{abc}$ and the integers $Y_{abc}$ (\ref{yfusion}). The importance of this second approach, 
although technically more complicated in the case at hand,  comes from the fact that it can be
used for essentially any left-right symmetric $2d$ CFT.

\section{Adding a Magnetic Field}

As in toroidal and orbifold compactifications of open strings, the N5-brane
backgrounds allow for the introduction  of a constant magnetic field. This corresponds to the
insertion on the boundary of an operator of the form \cite{acny}
\be {\cal B}^i = J^i + {i \over 2} \epsilon^i{}_{jk} \psi^j \psi^k
\label{magnet}
\ee 
This boundary deformation of the rational CFT is integrable and one can express the open-string spectrum in terms of the characters (\ref{character})
with $z$ related to the magnetic field and the charges of the open-string states $q_i$ by \cite{acny}
\be z = {1\over \pi} ( arctg(q_1 {\cal B}) + arctg(q_2 {\cal B}) )
\label{shift}
\ee 
Using the $SU(2)$ symmetry one can always choose  ${\cal B}$ pointing in
the third direction.
From the modular S-transformation \cite{kac}
\be
\chi_a^{(k)}(-{1\over\tau},-{z\over\tau},u-{z^2\over 2\tau}) = \sum_b S_{ab}
\chi_b^{(k)}(\tau,z,u)
\ee one immediately deduces the Casimir energy and the shift of the modes of the currents
$J^{(\pm )}_n \rightarrow J^{(\pm )}_{n\pm z}$. Notice that, since the modes of $J^{(3)}$ are
unaffected, the current algebra is preserved
\ba
 \left[ J^{(+)}_{n+z},J^{(-)}_{m-z} \right] &=& 2 J^{(3)}_{n+m} + k \delta_{n+m}
\nn
\\
\left[ J^{(3)}_{n},J^{(\pm )}_{m\pm z} \right] &=& \pm J^{(\pm )}_{n+m\pm z}
\\
 \left[ J^{(3)}_{n},J^{(3)}_{m} \right] &=& {k\over 2} \delta_{n+m}
\nn
\ea thus the introduction of the magnetic field simply amounts to a modulation of the boundary
reflection coefficients. By world-sheet supersymmetry considerations the modes of the  fermions
get an opposite shift. Indeed the total $N=1$ supercurrent, that couples to the worldsheet
gravitino,
\be G = J^i \psi_i + i \partial X^4 \psi_4 + {i \over 3!}\epsilon^{ijk} \psi_i \psi_j \psi_k + Q
\partial \psi_4
\ee 
 forbids a twist of $\psi_4$ (and similarly of $X^4$) due to the presence of the background
charge. The twisting of only two currents and two fermions leads to an explicit breaking of the
spacetime supersymmetry. Since the curvature of the spin connection with torsion is self-dual
one may ask if there is a possibility of adding a self-dual field-strength, \ie an instanton like
gauge field in order to make the background supersymmetric. A possibility of this kind is
suggested by the standard embedding in the heterotic version of the N5-brane. This issue
deserves further study and it may have interesting applications to other curved backgrounds such
as orbifolds, Gepner models and fermionic models. The final goal would be to address the issue of
consistency of magnetized D-branes inside Calabi-Yau manifolds \cite{hooy}.

\section{Conclusions}

In the curved spacetime  models at hand there are no tadpole conditions to be imposed, since
there are no massless states in the closed-string spectrum. Indeed, the contribution of the FF
boson shifts all masses by  $Q^2 / 8 = 1 / 4(k+2)$ thus all closed-string states become massive.
Moreover, the distinction between the different Dp-branes is smeared out for any finite  value
of the level $k$. In fact only the limit
$k
\rightarrow
\infty$ which corresponds to flat spacetime allows for a simple geometrical interpretation.

The systematic construction of open string models  elaborated in
\cite{bs}, further improved in \cite{bps,min} and finally completely established in
\cite{wzwb} has proven to be quite a powerful tool for  analyzing D-branes and O-planes in
non-trivial backgrounds, that is backgrounds that are not simply related by orbifolds to free
field theories and have non-abelian fusion rules.  Indeed the dynamics of the D-branes and
O-planes in the background of $k$ coincident N5-branes is almost completely captured by the open
string descendants of the $SU(2)$  WZNW models \cite{wzwa,wzwb}. The wormhole geometry however
is only the simplest instance of a large class of generalized hyper-K\"ahler manifolds that  due
to the enhanced
$N=(4,4)$ superconformal symmetry correspond to exact superstring backgrounds
\cite{kkl}. The open string
 descendants of the parent Type II superstring vacua seem at reach as well as the issues of
Buscher's T-dualities in this non-compact backgrounds with isometries \cite{bus}. 

Finally, much in the same way as in \cite{kk}, this curved background may be used as a consistent
string IR regulator in order to extract the stringy loop correction to the
low-energy effective Type I superstring lagrangian. For $N=1$ Type I vacua with D5-branes,
heterotic - Type I duality should map this corrections into non-perturbative corrections to the
heterotic string effective lagrangian. Elaborating a systematic approach to the study of
D-branes and their open-string excitations in non trivial backgrounds may thus provide a
powerful tool for investigating new $N=1$ dualities, that generalize the 
simple instances so far
considered both in the superstring and in the gauge theory setting.

\section{Acknowledgments} We would like to acknowledge fruitful discussions with C.~Angelantonj,
E.~Kiritsis, A.~Marshakov, D.~Polyakov, G.~Pradisi, A.~Sagnotti and M.~Tonin. Ya.~S. would like
to thank  the Physics Department at the University of Rome ``Tor Vergata" and I.N.F.N.  Sezione
di Roma2 for their hospitality  and financial support  while this work was done. Some
preliminary results of this work were presented by M.~B. at the ``Rome-Paris-Utrecht Triangular
Meeting" held  at Ecole Normale Superieure (Paris) in august and at the ``V Italo-Korean
Meeting" held at Sogang University (Seoul) in september.

\vfill
\eject

\renewcommand{\theequation}{\thesection.\arabic{equation}}

\appendix

\section{Appendix}

In this Appendix we collect some formulae 
concerning the open descendants of $SU(2)$ WZNW models
\cite{wzwa,wzwb}.

The central charge of the Virasoro algebra for the current algebra at 
level $k$ is
$c = 3k/(k+2)$ while the conformal weights of the integrable unitary representations  with
isospin $I$ in the range $I=0,..,k/2$ are
\be h_I^{(k)} = {I(I+1)\over k+2} .
\ee

The character formula is obtained from (\ref{character}) for $z=u=0$ 
\ba
\chi_I^{(k)}(\tau,0,0) = Tr_{{\cal H}_I^{(k)}} q^{L_o-{c\over 24}}  =    q^{h_I^{(k)}-{c(k)\over
24}} { \sum_{n} q^{(k+2)n^2+(2I+1)n} (2I+1+2 n(k+2)) 
\over  {\prod_{n=1}^{\infty} (1-q^n)^3}  } .
\ea

It is convenient to label states and characters in terms of the dimension $a=2I+1$ of the
corresponding highest weight representations of $SU(2)$.  The modular matrices in the above
basis are  \cite{kac}
\be S_{ab} \  = \  \sqrt{{2 \over k+2}}  \ sin \left({\pi a b  \over k+2}\right) \ ,
\label{smatrix}
\ee and
\be T_{ab} \  =  \ \delta_{a b} \  e^{i \pi \left( {a^2 \over 2(k+2)} - {1 \over 4}\right)}
\quad .
\label{tmatrix}
\ee The charge conjugation matrix is equal to the identity  $C=S^2 = (ST)^3 = 1$.  The modular
transformation between the direct and the transverse 
channel of the M\"obius strip is induced by 
$P = T^{1/2} S T^2 S T^{1/2}$ that acts on {\it hatted} characters \cite{bs}
\be
\hat\chi_h (i\tau_2 + 1/2)  = e^{-i \pi (h-c/24)} \chi_h (i\tau_2 + 1/2)  = T^{-1/2} \chi_h
(i\tau_2 + 1/2)
\ee
 and in general satisfies 
$P^2 = C$. For the $SU(2)$ WZNW model $P$ is represented by
\be P_{ab} \  =  \ {2 \over \sqrt{k+2}} \  sin \left( {\pi a b  \over 2(k+2)}\right)
 \ (E_k  E_{a+b} + O_k O_{a+b}) \ ,
\label{pmatrix}
\ee where $E_n$ and $O_n$ are projectors on even and odd $n$  respectively. The fusion rule
coefficients are given by the Verlinde formula \cite{ver}
\be {N_{ab}}^c \  =  \ \sum_{d=1}^{k+1} \ {S_{ad} S_{bd} S^{\dagger}_{cd} \over S_{1d}} \ = \ 
\left\{  { 1 \ {\rm if} \ |a-b|+1 \leq c \leq min(a+b-1,2k-a-b+3)
\atop 0 \ {\rm else} \hfill}
\right.
\quad ,
\label{nfusion}
\ee  It turns out to be convenient to introduce also the integer (!) coefficients
\cite{wzwa} 
\be {Y_{ab}}^c  \ =  \ \sum_{d=1}^{k+1}  \ {S_{ad} P_{bd} P^{\dagger}_{cd} \over S_{1d}}
\quad .
\label{yfusion}
\ee 

Cappelli, Itzykson and Zuber \cite{ciz} have shown that the modular invariant torus partition
functions are  in one to one correspondence with the $A-D-E$ simply laced simple Lie algebras.
At any level $k$ there is  a diagonal modular invariant  denoted by $A_{k+1}$ that reads
\be Z^{\{A_{k+1}\}} \ =  \ \sum_{a=1}^{k+1} | \chi_a |^2 \quad .
\label{aseries}
\ee For $k=4p+2$, there is also a permutation modular invariant, denoted by $D_{2p+1}$, 
\be Z^{\{D_{2p+1}\}}  \ =  \ \sum_{{\rm odd} \  a=1}^{4p-1} |\chi_a|^2 + |\chi_{2p+2}|^2 +
\sum_{{\rm even} \  a=2}^{2p} (\bar \chi_a \chi_{4p+4-a} +
\bar \chi_{4p+4-a} \chi_a)
\quad .
\label{dodseries}
\ee The 
$D_{even}$ series (present for level $k=4p$) and the three $E$ cases correspond to extended 
chiral algebras. 
 
The unoriented projection can be chosen to preserve the diagonal $SU(2)$ subalgebra of the
$SU(2)_L \times SU(2)_R$ current algebra symmetry. Corresponding to the two geometrical
involution on the $SU(2)$ group manifold, \ie
$g\rightarrow -g^{-1}$ and $g\rightarrow g^{-1}$, there are two different unoriented projections 
(Klein bottle amplitudes) of the parent torus partition function \cite{hor}. The
action of the two involutions coincides on the integer isospin representations, while on  the half-integer isospin representations the first one corresponds to keeping  the
singlets of the diagonal $SU(2)$, and the second one corresponds to 
removing them from the spectrum.

We shall label the two choices  by an index $r$ and
$c$ in order to streamline their relation to real and complex CP charge assignments. The diagonal
$A_{k+1}$ models allow for the introduction of 
$k+1$ CP multiplicities or equivalently $k+1$ independent boundary states  that are in one to
one correspondence with the integrable $SU(2)$ representations. For the $D_{2p+1}$ models there
is a reduction of the independent boundary states and only $k/2+2$ CP charges can be introduced
\cite{wzwb}.

For the $A$-series  with real CP charges the Klein (K), annulus (A) and M\"obius strip (M) direct
(``loop") channel amplitudes read \cite{wzwa}
\ba
  K_r^{\{A_{k+1}\}} 
\ &=& \ \sum_{a=1}^{k+1} \  {Y^a}_{11}  \chi_a \ = \ \sum_{a=1}^{k+1} (-1)^{a-1} \chi_a \quad ,
\\ A_r^{\{A_{k+1}\}} \ &=& \sum_{a,b,c=1}^{k+1} {N_{ab}}^c \chi_c n^a n^b\quad ,   \\
M_r^{\{A_{k+1}\}} \ &=& \ \pm \sum_{a,b=1}^{k+1} {Y_{a1}}^{b} \hat \chi_b n^a 
\ = \ \pm \sum_{a,b=1}^{k+1} (-1)^{a-1} (-1)^{b-1 \over 2}  {N_{aa}}^b \hat \chi_b n^a \quad .
\label{areald}
\ea A modular transformation yields the transverse (``tree") channel amplitudes
\ba
\tilde K_r^{\{A_{k+1}\}} \ &=& \ \sum_a \left( {P_{1a} \over
\sqrt{S_{1a}} } \right)^2 \chi_a \quad , \\
\tilde A_r^{\{A_{k+1}\}} \ &=&  \ \sum_a \left(  
\sum_b { S_{ab} n^b \over \sqrt{S_{1a}}}
\right)^2 \chi_a \quad , \\
\tilde M_r^{\{A_{k+1}\}} &=& \pm 
\sum_a \left(  
\sum_b { P_{1a} S_{ab} n^b \over S_{1a}} \right) \hat \chi_a \quad .
\label{arealt}
\ea

For the $A$-series with complex CP charges the various the direct channel
partition functions read
\ba K_c^{\{A_{k+1}\}} \ &=&   \ \sum_{a=1}^{k+1} \  {Y^a}_{k+1,k+1}  \chi_a 
\ = \ \sum_{a=1}^{k+1} \chi_a \quad , 
\\ A_c^{\{A_{k+1}\}} \ &=&  \ \sum_{a,b,d=1}^{k+1} {N_{ab}}^d\chi_{k+2-d}  n^a n^b\quad , 
\\ M_c^{\{A_{k+1}\}} \ &=&  \ \pm \sum_{a,b=1}^{k+1} {Y_{a,k+1}}^{b} \hat \chi_b n^a 
\ = \ \pm \sum_{a,b=1}^{k+1}  {N_{aa}}^b \hat \chi_{k+2-b} n^a \quad . 
\label{acomplexd}
\ea The transverse channel amplitudes are then given by 
\ba
\tilde K_c^{\{A_{k+1}\}} \ &=& \ \sum_a \left( {P_{k+1,a} \over
\sqrt{S_{1a}} } \right)^2 \chi_a  \quad , 
\\
\tilde A_c^{\{A_{k+1}\}} \ &=&  \ \sum_a (-1)^{a-1} \ \left(  
\sum_b { S_{ab} n^b \over \sqrt{S_{1a}}} \right)^2 \chi_a \quad , \\
\tilde M_c^{\{A_{k+1}\}} \ &=&  \ \pm \sum_a \left( \sum_b { P_{k+1,a} S_{ab} n^b \over S_{1a}}
\right) \hat \chi_a  \quad .
\label{acomplext}
\ea Notice that positivity of the transverse channel requires the numerical identifications
$n_{k+2-a} = \bar n_a = n_a$.

For the $D_{2p+1}$-series, at level $k=4p+2$, the two choices for the Klein bottle are
\cite{wzwb}
\be K_r^{\{D_{2p+1}\}} = \left( 
\sum_{{\rm odd} \ a=1}^{k+1}  \chi_a - \chi_{k/2+1} \right) \quad , 
\label{kdodreald}
\ee and
\be K_c^{\{D_{2p+1}\}} = \left( 
\sum_{{\rm odd} \ a=1}^{k+1}  \chi_a + \chi_{k/2+1} \right) \quad .
\label{kdodcomplexd}
\ee Correspondingly, in the transverse channel one finds
\be
\tilde K_r^{\{D_{2p+1}\}}\ = \ \sum_{a=1}^{k+1} \left( (-1)^{{ a^2-1 \over
8}}{P_{k/2,a}\over\sqrt{S_{1a}} } \right)^2 \chi_a \quad , 
\label{kdodrealt}
\ee and
\be
\tilde K_c^{\{D_{2p+1}\}} \ = \ \sum_{a=1}^{k+1} \left( (-1)^{{ a^2-1 \over
8}}{P_{k/2+2,a}\over\sqrt{S_{1a}} } \right)^2 \chi_a \quad .
\label{kdodcomplext}
\ee

The annulus and M\"obius strip partition functions can be  expressed either  in terms of $k+1$
{\it linearly dependent} pseudo-charges $\nu_a$ as in \cite{wzwb} or can be read off the
expressions presented in Section 4.

\end{document}